# Long spin coherence length and bulk-like spin-orbit torque in ferrimagnetic multilayers

Jiawei Yu[1], Do Bang[2†], Rahul Mishra[1†], Rajagopalan Ramaswamy[1], Jung Hyun Oh[3], Hyeon-Jong Park[4], Yunboo Jeong[5], Pham Van Thach[2], Dong-Kyu Lee[3], Gyungchoon Go[3], Seo-Won Lee[3], Yi Wang[1], Shuyuan Shi[1], Xuepeng Qiu[6], Hiroyuki Awano[2], Kyung-Jin Lee[3,4,5*], and Hyunsoo Yang[1*]

[1] Department of Electrical and Computer Engineering, National University of Singapore, 117576, Singapore

[2] Toyota Technological Institute, Tempaku, Nagoya 468-8511, Japan

[3] Department of Materials Science and Engineering, Korea University, Seoul 02841, Korea

[4] KU-KIST Graduate School of Conversing Science and Technology, Korea University, Seoul 02841, Korea

[5] Department of Semiconductor Systems Engineering, Korea University, Seoul 02841, Korea

[6] Shanghai Key Laboratory of Special Artificial Macrostructure Materials and Technology and School of Physics Science and Engineering, Tongji University, Shanghai 200092, China

†These authors contributed equally to this work.

*e-mail: eleyang@nus.edu.sg, kj_lee@korea.ac.kr

**Spintronics is a multidisciplinary field whose central theme is the active manipulation of spin degrees of freedom in solid-state systems. Ferromagnetic spintronics has been a main focus as it offers non-volatile memory and logic applications through current-induced spin-transfer torques[1-4]. Enabling wider applications of such magnetic devices requires a lower switching current for a smaller cell while keeping the thermal stability of magnetic cells for non-volatility. As the cell size reduces, however, it becomes extremely difficult to meet this requirement with ferromagnets because spin-transfer torque for ferromagnets is a surface torque due to rapid spin dephasing[5,6], leading to the 1/ferromagnet-thickness dependence of the spin-torque efficiency[7]. Requirement of a larger switching current for a thicker and thus**

**more thermally stable ferromagnetic cell is the fundamental obstacle for high-density non-volatile applications with ferromagnets. Theories predicted that antiferromagnets have a long spin coherence length due to the staggered spin order on an atomic scale[8,9], thereby resolving the above fundamental limitation. Despite several spin-torque experiments on antiferromagnets[10-12] and ferrimagnetic alloys[13-16], this prediction has remained unexplored. Here we report a long spin coherence length and associated bulk-like-torque characteristic in an antiferromagnetically coupled ferrimagnetic multilayer. We find that a transverse spin current can pass through > 10 nm-thick ferrimagnetic Co/Tb multilayers whereas it is entirely absorbed by 1 nm-thick ferromagnetic Co/Ni multilayer. We also find that the switching efficiency of Co/Tb multilayers partially reflects a bulk-like-torque characteristic as it increases with the ferrimagnet-thickness up to 8 nm and then decreases, in clear contrast to 1/thickness-dependence of Co/Ni multilayers. Our results on antiferromagnetically coupled systems will invigorate researches towards energy-efficient spintronic technologies.**

The spin-transfer torque (STT) acting on ferromagnets (FMs) is a surface torque, based on the averaging effect of STT[5,6]. We note that the same averaging effect occurs regardless of the spin-current source, and the spin-orbit torque (SOT)[17,18], which we use in our experiment, is also a surface torque for FMs (Extended Data Fig. 1 and Methods). When a transverse spin current with a spin orientation non-collinear with the magnetization is injected into a FM, the electron spin precesses rapidly in real space because the wave vectors of the majority (↑) and minority (↓) spins at the Fermi surface are different (i.e., $k_F^\uparrow \neq k_F^\downarrow$). The precession wavelengths are different for different incident angles of electrons (i.e., the direction of wave vector *k*), leading to rapid spin dephasing when summing over all current-carrying *k*-states. As a result, the *k*-integrated transverse spin current decays to zero within a distance from the FM surface, called the ferromagnetic coherence length (spin coherence length, more generally), $\lambda_c = \pi / |k_F^\uparrow - k_F^\downarrow|$.[19] As $|k_F^\uparrow - k_F^\downarrow|$ becomes larger for larger exchange splitting, $\lambda_c$ is only a few angstroms in strong FMs (e.g. cobalt or iron) for which the STT is almost a surface torque.

Theories predicted that the spin coherence length is very long in antiferromagnets (AFMs) because of the staggered spin order on an atomic scale[8,9]. We use the term of "bulk-like-torque" to describe the characteristic of spin-torque for AFMs, i.e., spin-current absorption on a larger thickness, in contrast to the surface-torque of FMs. A semi-classical explanation of bulk-like-torque is that for conduction electron spins, the moments with alternating orientation on an atomic scale are seen as the exchange interactions with alternating signs. As a result, an ideal AFM has zero net effective exchange interaction when averaged over two sub-lattices and thus has an infinitely long $\lambda_c$, yielding the bulk-like-torque characteristic. Several experiments have investigated on STT/SOT effects in systems including AFMs[10-12] and more recently on ferrimagnetic alloys[13-16], but not on the long spin coherence length and associated bulk-like-torque characteristic.

We qualitatively illustrate the spin coherence length in FMs and FIMs (or AFMs) based on the spin precession around the local exchange field. Neglecting the spin relaxation, dynamics of non-equilibrium spin density **s** is described by $\partial \mathbf{s}/\partial t = -\gamma\, \mathbf{s} \times \mathbf{H}_{ex}$, where $\gamma$ is the gyromagnetic ratio, $\mathbf{H}_{ex}$ is the effective exchange field that is aligned along the local magnetic moment **m**. Assuming $\mathbf{s} = (\sin\theta, \cos\theta, 0)$ and $\mathbf{H}_{ex} = H\hat{\mathbf{z}}$, this equation of motion transforms to $\partial\theta/\partial t = \delta\theta/\delta t = -\gamma H$, where the sign of spin precession angle $\delta\theta$ follows the sign of $H$ and thus the sign of **m** (Fig. 1a and b). In a FM, an electron spin propagating along the *x*-direction continuously precesses in the same sense because of the homogeneous exchange field (Fig. 1c). On the other hand, in a FIM, an electron spin precesses counter-clockwise on a lattice corresponding to a positive exchange field, whereas it precesses clockwise on the next lattice corresponding to a negative exchange field. As a result, the period (or wavelength) of spin precession in FIMs is longer than that in FMs, resulting in much less spin dephasing.

In order to verify the theoretical prediction of long spin coherence length, we perform experiments with a ferrimagnet (FIM), i.e., Co/Tb multilayers where both Co and Tb layers are atomically thin and their moments are coupled antiferromagnetically. We choose a FIM, instead of an AFM, for following two reasons. One is that Co/Tb multilayers can show a longer $\lambda_c$ than a FM because of the antiferromagnetic alignment of Co and Tb moments (Extended Data Fig. 1 and Methods), thereby exhibiting a feature of the bulk-like-torque characteristic. As explained above, the STT efficiency of a FM is inversely proportional to the FM-thickness whereas that of an ideal AFM is independent of the AFM-thickness. As $\lambda_c$ of FIM is located between those of FM and AFM, it is expected that the STT efficiency of a FIM first increases and then decreases with the FIM-thickness. The other reason to choose FIMs is that various measurement methods established for FMs are applicable to FIMs because of nonzero net moment[13,20]. However, the choice of FIM also results in a difficulty. FIMs commonly show a thickness-dependent variation of magnetic properties[21], as also observed in Co/Tb multilayers (Extended Data Fig. 4 and Methods), which makes a quantitative analysis of spin transport difficult. Even with this difficulty, our thickness-dependent SOT measurements combined with spin pumping measurements support a long spin coherence length and associated bulk-like-torque characteristic in ferrimagnetic Co/Tb multilayers, as we show below.

We fabricate perpendicularly magnetized ferromagnetic [Co/Ni]$_N$ and ferrimagnetic [Co/Tb]$_N$ multilayers (Fig. 2a, b; see Methods for details), where the total thickness varies with changing the repetition number $N$. Both Co/Ni and Co/Tb multilayers have an additional Pt layer, and an in-plane current generates SOTs. We use the harmonic Hall voltage measurements to quantify the strength of SOT effective fields[22,23]. The longitudinal and transverse measurement schematics are illustrated in Fig. 2c and d, respectively. Representative results for the longitudinal (blue line) and transverse (red line) second harmonic voltages ($V_{2f}$) from Co/Ni ($N = 2$) and Co/Tb ($N = 5$) devices are shown in Fig. 2e and f, respectively. The anomalous Nernst effect is corrected in the $V_{2f}$ data[22]. We observe a clear $V_{2f}$ in the longitudinal configuration (H//I), which is mostly determined by the anti-damping SOT[22,23]. The opposite $V_{2f}$ signs in the H//I case for Co/Ni and Co/Tb multilayers indicate that the Pt layer is the source of spin currents, as it is placed on top of the Co/Tb multilayer, but under the Co/Ni multilayer. In order to rule out the contribution from pure bulk Co/Tb to SOT, we conduct a control experiment without and with the spin current source, Pt (Extended Data Fig. 5, 6 and Methods). We find that there is no noticeable current-induced SOT without the Pt layer, suggesting that the Co/Tb bulk itself cannot directly contribute to SOTs.

We extract the spin-orbit effective fields, $H_L$ and $H_T$, by fitting $V_{2f}$,[23] where $H_L$ and $H_T$ correspond to the anti-damping (longitudinal) and field-like (transverse) components of SOTs, respectively. The planar Hall effect is considered for the fitting (Extended Data Fig. 8 and Methods). Devices with different $N$, corresponding to different thicknesses, $t_{FM}$ or $t_{FIM}$, have been measured. Absolute SOT effective fields normalized by the current density in the Pt layer ($H_{L/T}/J$) are presented in Fig. 3a and b for the Co/Ni and Co/Tb systems, respectively. We find that both $H_L$ and $H_T$ of Co/Ni multilayers decrease as $t_{FM}$ increases, consistent with the surface-torque characteristic expected for FMs. However, Co/Tb multilayers show an entirely different trend, in which both $H_L$ and $H_T$ increase up to $t_{FIM}$ of 7.9 nm and then decrease for thicker samples.

We estimate the effective spin Hall angle $\theta_{eff} = H_L(2eM_S t_{FM/FIM}/\hbar J)$, where $e$ is the electron charge and $\hbar$ is the reduced Planck's constant, with considering thickness-dependent variation of the saturation magnetization $M_S$ and current density $J$ in the Pt layer. The Co/Tb multilayer shows a significant $M_S$ variation with the minimum at $t_{FIM}$ of 6.6 nm (inset of Fig. 3d). We find that $\theta_{eff}$ of Co/Ni multilayer is nearly constant with $t_{FM}$ (Fig. 3c). Similar to the tendency of $H_L/J$, $\theta_{eff}$ of Co/Tb multilayer increases up to $t_{FIM}$ = 9.9 nm and then decreases for a thicker sample (Fig. 3d). Besides the distinct thickness-dependence of $\theta_{eff}$, another interesting observation is that the Co/Tb multilayer shows a larger $\theta_{eff}$ than Co/Ni multilayer ($\theta_{eff}$ of Co/Tb multilayer = 2.1 at $t_{FIM}$ of 9.9 nm and the average $\theta_{eff}$ of Co/Ni multilayer = 0.2 ± 0.05). We note that a model calculation with considering a thickness-dependent variation of the *sd* exchange in FIMs shows qualitatively similar trends with the experimental ones (Extended Data Fig. 2 and Methods), even though the model is too simple to capture all the details of FIMs. Nevertheless, this qualitative agreement between model and experiment results indicates that the distinct behavior of $\theta_{eff}$ of Co/Tb multilayer would originate from a combined effect of long spin coherence length and thickness-dependent property variation.

As an independent test, we perform SOT switching experiments with applying an external field ($H_{ext}$) in the current direction ($\theta$ = 0°) for deterministic switching. The insets of Fig. 3e and f show the representative current-induced switching data obtained from Co/Ni ($N$ = 2) and Co/Tb ($N$ = 5) samples, respectively. As the switching is governed by domain nucleation and propagation in large samples (i.e., Hall bar width = 10 μm), we estimate the STT efficiency $\eta = H_p / J$, where $H_p$ is the domain wall depinning field[24]. Figure 3e and f show $\eta$ as a function of $t_{FM}$ and $t_{FIM}$, respectively. For Co/Ni multilayers, $\eta$ decreases with $t_{FM}$, whereas for Co/Tb multilayers it increases and then decreases with $t_{FIM}$, following similar trends to SOT effective fields (Fig. 3a and b). We note that recently reported fast dynamics at the angular momentum compensation condition in FIMs[20] would affect the switching data, but not the harmonic Hall data.

As different approaches for the estimation of spin torque efficiency show qualitatively similar trends, it indicates that SOT for Co/Tb multilayers is not a surface torque. Moreover, the observed thickness-dependence of spin torque efficiency is qualitatively consistent with the model calculation (Extended Data Fig. 2 and Methods) for the bulk-like-torque characteristic in FIMs; it first increases and then decreases with the FIM-thickness. However, because of the thickness-dependent property variations in Co/Tb multilayers (inset of Fig. 3d and Methods), this result is not yet conclusive but it is still possible that another unknown mechanism is responsible for the distinct thickness-dependence observed in Co/Tb multilayers.

In order to resolve this ambiguity, we perform additional spin pumping experiments to estimate the spin coherence length $\lambda_c$. We measure a spin-pumping-induced inverse spin Hall voltage ($V_{ISHE}$) for substrate/Pt(10)/[FIM or FM]/Cu(2.4)/Co(20) structures (numbers in nanometers; FIM = [Co(0.32)/Tb(0.34)]$_N$ and FM = [Co(0.3)/Ni(0.6)]$_N$) as shown in Fig. 4a. In these structures, the Co/Ni and Co/Tb multilayers are perpendicularly magnetized, whereas the top thick Co layer has an in-plane magnetization. In the spin pumping setup (Fig. 4b; see Methods for details), the top Co layer generates a spin-pumping-induced spin current with an in-plane spin polarization (thus transverse to Co/Ni or Co/Tb magnetization direction), which passes through the Cu layer and enters the Co/Ni or Co/Tb layer. If $\lambda_c$ of the Co/Tb multilayer is long, it is expected that a transverse spin current passes through the Co/Tb layer without much spin dephasing and reaches the bottom spin sink, Pt, and subsequently, $V_{ISHE}$ is generated by the inverse spin Hall effect of Pt. On the other hand, $V_{ISHE}$ is expected to be negligible for a thick Co/Ni multilayer because a transverse spin current is almost absorbed at the [Co/Ni]/Cu interface. Therefore, the measurement of $V_{ISHE}$ versus FIM- or FM-thickness provides an estimate of $\lambda_c$.

We find that the experimental results are consistent with this expectation. In Fig. 4c, black symbols are the data from a reference Pt/Cu/Co sample, which shows the largest $V_{ISHE}$ signal at an in-plane bias field $H_b$. For the Co/Ni-based structure, $V_{ISHE}$ signal becomes negligible at a Co/Ni thickness of 0.9 nm (blue symbols). In contrast, $V_{ISHE}$ signal for the Co/Tb-based structure is finite at a much thicker Co/Tb (red symbols, Co/Tb thickness = 5.3 nm as an example). $V_{ISHE}$ signal disappears when excluding the top Co layer from the Co/Tb-based structure (green symbol), proving that the perpendicularly magnetized Co/Tb itself does not generate a $V_{ISHE}$ signal. In Fig. 4d, spin pumping results are summarized for a wide thickness range of Co/Tb multilayer. It shows that $V_{ISHE}$ signal is finite even at 13 nm-thick Co/Tb. This result evidences a long spin coherence length in ferrimagnetic Co/Tb multilayers.

The spin pumping and spin torque are connected through the Onsager reciprocity[25]. Therefore, the long spin coherence length observed in spin pumping experiments suggests that the bulk-like-torque characteristic must be present in spin torque experiments at least partially. Given that the thickness-dependent change in the spin torque efficiency follows the trend expected for the bulk-like-torque characteristic (Fig. 3), the spin pumping experiments combined with the spin torque experiments allow us to conclude that the antiferromagnetically coupled FIMs show a long spin coherence length and associated bulk-like-torque characteristic. We note that this bulk-like-torque characteristic and equivalently long spin coherence length are also observed for FIM alloys (Extended Data Fig. 9 and Methods), which would relate to some ordering in FIM alloys[26,27]. The results were supported by model calculation using a ferrimagnetic alloy (Extended Data Fig. 2 and Methods) in which an ordered alloy shows a longer spin coherence length than a random alloy. These salient features make antiferromagnetically coupled FIMs attractive for low-power non-volatile applications. We expect the bulk-like-torque principle is also applicable to domain wall or skyrmion devices[28-30] operated by SOTs. In this respect, our findings will motivate research activities to introduce FIMs as core elements in spintronics devices, which have been so far dominated by FMs. Therefore, our result provides an important step towards "ferrimagnetic spintronics".

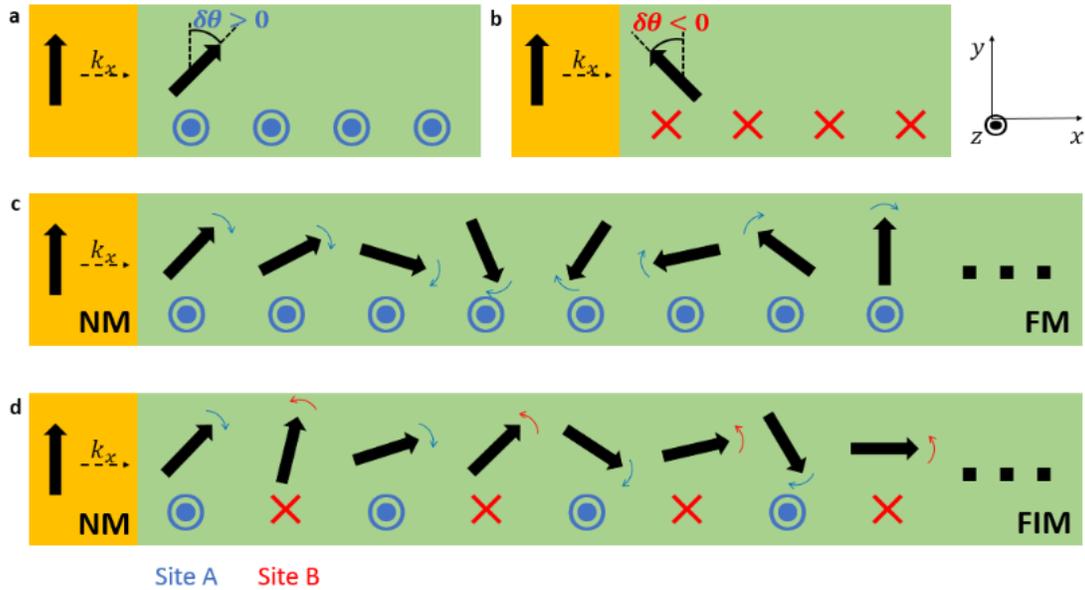

**Figure 1 | Schematic illustrations of spin precession from the semi-classical viewpoint. a, b,** Local spin precession angle $\delta\theta$ in a FM with up magnetic moment ($\mathbf{m}//+\hat{\mathbf{z}}$) and down magnetic moment ($\mathbf{m}//-\hat{\mathbf{z}}$), respectively. Blue dots (Fig. 1a) and red crosses (Fig. 1b) indicate the directions of magnetic moments. Precession of an electron spin in the FM layer (**c**) and in the FIM layer (**d**). Blue and red curved arrows indicate $\delta\theta > 0$ and $\delta\theta < 0$, respectively.

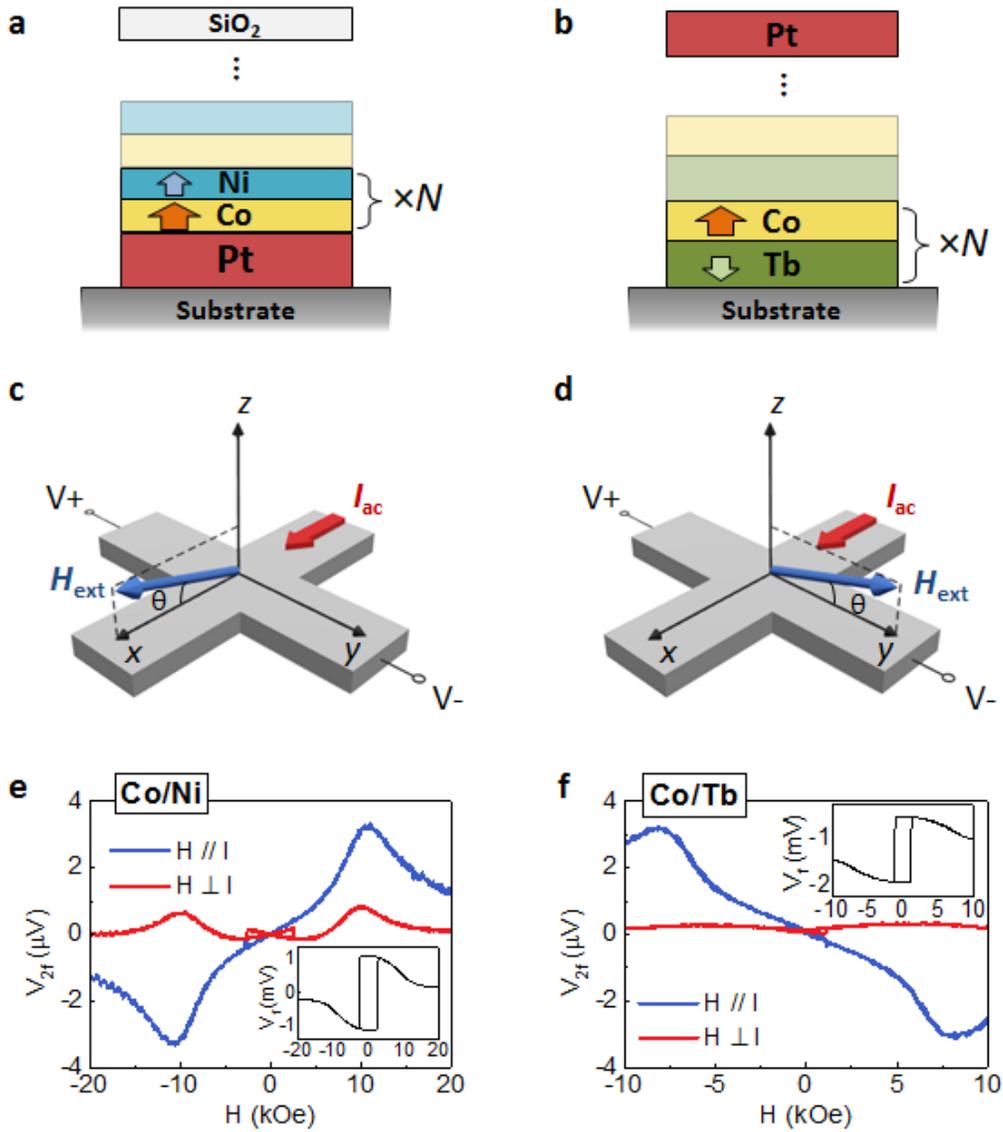

**Figure 2 | Film stacks and SOT measurements. a, b,** Illustrations of Co/Ni (**a**) and Co/Tb (**b**) multilayers. The magnetizations of Co, Ni and Tb sub-lattices are presented by the yellow, blue and green arrows, respectively. **c, d,** The measurement schematics for longitudinal (**c**) and transverse (**d**) SOT effective fields. **e, f,** Second harmonic voltages ($V_{2f}$) obtained from Co/Ni (**e**) and Co/Tb (**f**) multilayer devices, with the blue curves representing the longitudinal signals and red curves representing the transverse signals. The insets correspond to first harmonic voltages ($V_f$).

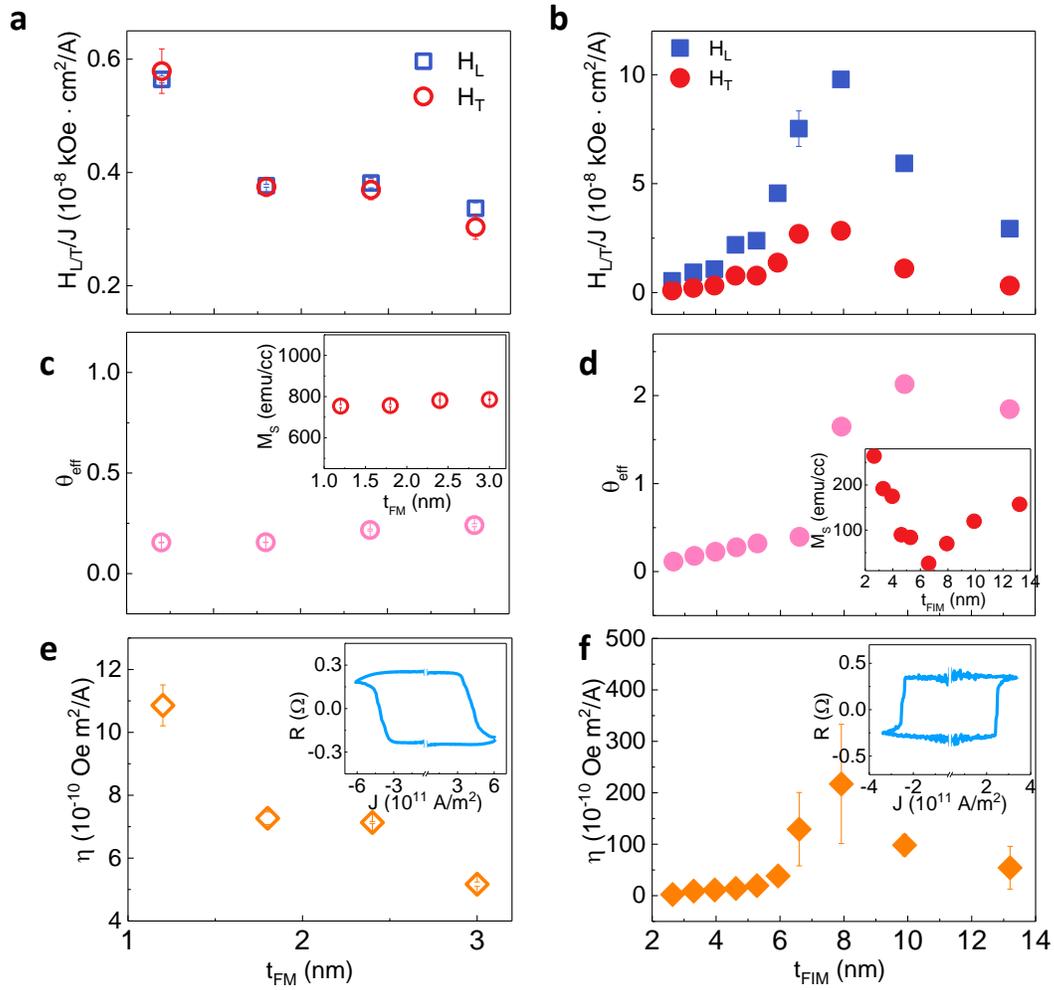

**Figure 3 | SOT effective fields and switching efficiencies. a, b,** Longitudinal ($H_L$) and transverse ($H_T$) SOT effective fields as a function of Co/Ni (**a**) or Co/Tb (**b**) thicknesses. **c, d,** Effective spin Hall angle ($\theta_{eff}$) as a function of Co/Ni (**c**) or Co/Tb (**d**) thicknesses. Insets in **c** and **d** are the saturation magnetization ($M_S$) as a function of ferromagnet- or ferrimagnet-thickness. **e, f,** Switching efficiencies ($\eta$) as a function of Co/Ni (**e**) or Co/Tb (**f**) thicknesses. Insets in **e** and **f** are current-induced switching data, showing the Hall resistance ($R_H$) as a function of applied pulse current.

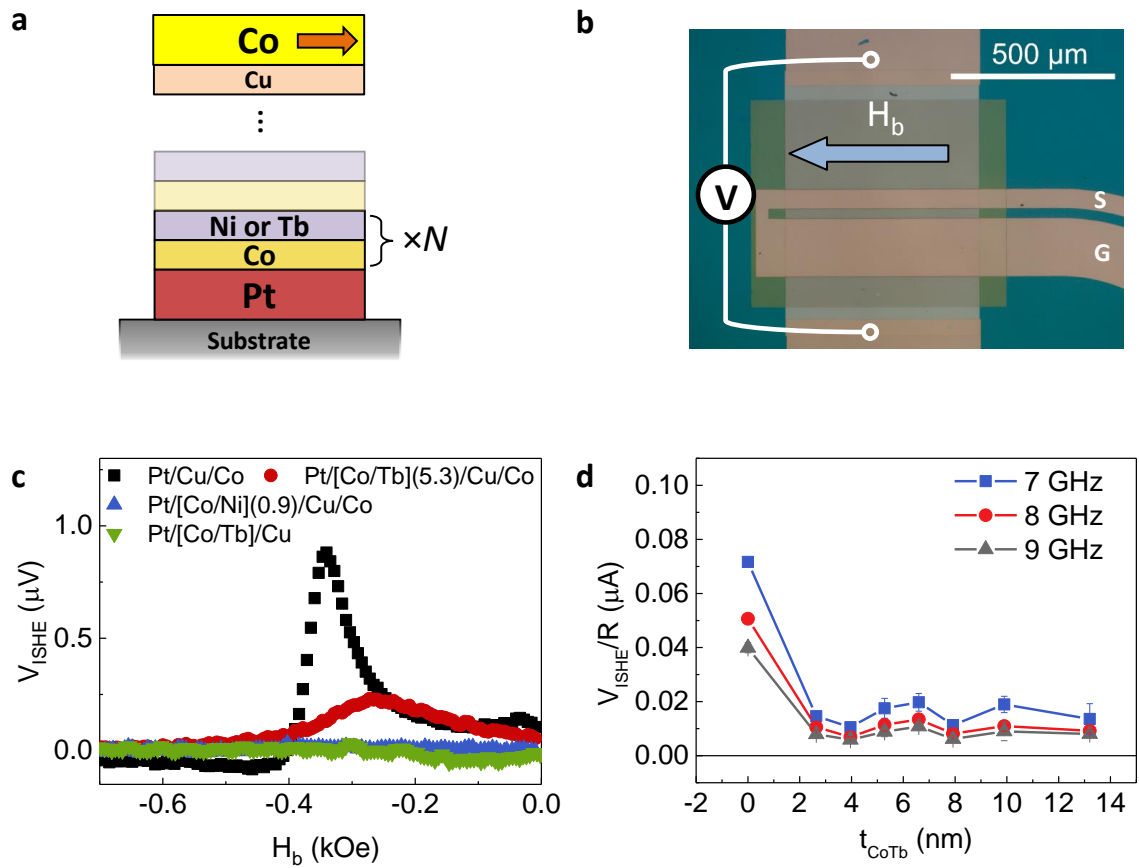

**Figure 4 | Spin pumping measurements. a,** Spin pumping sample structure. **b,** Schematic of spin pumping measurements. S and G indicates signal and ground connection for high frequency measurements. An in-plane field ($H_b$) along the waveguide direction is applied. **c,** Spin pumping signals in various structures. **d,** Inverse spin Hall signal as a function of [Co/Tb]-thickness in Pt/[Co/Tb]/Cu/Co structures at various frequencies.

## Methods

**Sample preparation**

Substrate/[Tb (0.34 nm)/Co (0.32 nm)]$_N$/Pt (4 nm) and substrate/MgO (2 nm)/Pt (4 nm)/[Co (0.3 nm)/Ni (0.3 nm)]$_N$/SiO$_2$ (3 nm) multilayers are fabricated on thermally oxidized silicon substrates using rf and dc magnetron sputtering system with a base pressure of ~ 10$^{-9}$ Torr. $N$ is the repetition number of Tb/Co or Co/Ni bilayer pairs, which is varied from 4 to 20 for Co/Tb systems and from 2 to 5 for Co/Ni systems. For Co/Tb multilayers, a 4 nm-thick Pt layer is deposited on top as a spin current source which also protects the multilayer from being oxidized. For Co/Ni multilayers, a bilayer of MgO (2 nm)/Pt (4 nm) is deposited on the bottom as a buffer and spin current source, and a SiO$_2$ (3 nm) layer is deposited as a capping layer to prevent possible oxidation of the FM layer. Subsequent photolithography and ion milling processes are performed to fabricate the films into Hall bar devices.

**Second harmonic and spin pumping measurements**

For the second harmonic measurements, an ac current $I_{ac}$ with a frequency of 13.7 Hz and a magnitude of 5 mA is injected into the channel of the device[31]. An external magnetic field $H_{ext}$ is applied along (orthogonal to) the current direction with a small out-of-plane tilting of $\theta = 4°$ from the film plane in the longitudinal (transverse) configuration. The first and second harmonic Hall voltages are recorded simultaneously by using two lock-in amplifiers triggered at the same frequency by the current source.

In the spin pumping measurements, a microwave at 7 to 9 GHz is applied to the asymmetric coplanar stripline waveguide by a signal generator. An in-plane field ($H_b$) along the waveguide direction is swept around the resonance field ($H_0$) given by the Kittel formula $f = \frac{\gamma}{2\pi}\sqrt{H_0(H_0 + 4\pi M_S)}$, where $\gamma$ is the gyromagnetic ratio and $M_S$ is the saturation magnetization. The voltage ($V$) is recorded by a lock-in amplifier. $V$ includes the asymmetric component ($V_{asym}$) from the anomalous Hall effect (AHE) and the anisotropic magnetoresistance, as well as the symmetric components ($V_{sym}$) from the spin pumping induced inverse spin Hall voltage ($V_{ISHE}$). Thus the measured voltage is fitted by a sum of symmetric and asymmetric Lorentzian function $V = V_{sym} \dfrac{\Gamma^2}{\Gamma^2 + (H-H_0)^2} + V_{asym} \dfrac{\Gamma(H-H_0)}{\Gamma^2 + (H-H_0)^2}$, from which $V_{ISHE}$ is extracted as $V_{sym}$.